\documentclass[11pt,twoside,a4paper]{article}

\usepackage{url}
\usepackage{ifthen}                 
\usepackage{fancyhdr}               
\usepackage{longtable}              
\usepackage{afterpage}              
\usepackage[footnotesize,centerlast]{caption}
\usepackage[round,sort]{natbib}       
\bibpunct{[}{]}{,}{n}{}{;}     
\bibpunct{[}{]}{,}{a}{}{;}
\usepackage[nottoc]{tocnatbibind}   
\makeatletter
\newenvironment{proof}{\vskip 2mm\noindent {\it Proof}:}{\hfill $\square$ \vskip 2mm} 

\renewcommand\section{\@startsection{section}{1}{\z@}%
                       {-18\p@ \@plus -4\p@ \@minus -4\p@}%
                       {12\p@ \@plus 4\p@ \@minus 4\p@}%
                       {\normalfont\large\bfseries\boldmath
                        \rightskip=\z@ \@plus 8em\pretolerance=10000 }}
\renewcommand\subsection{\@startsection{subsection}{2}{\z@}%
                       {-18\p@ \@plus -4\p@ \@minus -4\p@}%
                       {8\p@ \@plus 4\p@ \@minus 4\p@}%
                       {\normalfont\normalsize\bfseries\boldmath
                        \rightskip=\z@ \@plus 8em\pretolerance=10000 }}
\renewcommand\subsubsection{\@startsection{subsubsection}{3}{\z@}%
                       {-18\p@ \@plus -4\p@ \@minus -4\p@}%
                       {-0.5em \@plus -0.22em \@minus -0.1em}%
                       {\normalfont\normalsize\bfseries\boldmath}}
\renewcommand\paragraph{\@startsection{paragraph}{4}{\z@}%
                       {-12\p@ \@plus -4\p@ \@minus -4\p@}%
                       {-0.5em \@plus -0.22em \@minus -0.1em}%
                       {\normalfont\normalsize\itshape}}
\renewcommand\subparagraph[1]{\typeout{LLNCS warning: You should not use
                  \string\subparagraph\space with this class}\vskip0.5cm
You should not use \verb|\subparagraph| with this class.\vskip0.5cm}

\DeclareMathSymbol{\Gamma}{\mathalpha}{letters}{"00}
\DeclareMathSymbol{\Delta}{\mathalpha}{letters}{"01}
\DeclareMathSymbol{\Theta}{\mathalpha}{letters}{"02}
\DeclareMathSymbol{\Lambda}{\mathalpha}{letters}{"03}
\DeclareMathSymbol{\Xi}{\mathalpha}{letters}{"04}
\DeclareMathSymbol{\Pi}{\mathalpha}{letters}{"05}
\DeclareMathSymbol{\Sigma}{\mathalpha}{letters}{"06}
\DeclareMathSymbol{\Upsilon}{\mathalpha}{letters}{"07}
\DeclareMathSymbol{\Phi}{\mathalpha}{letters}{"08}
\DeclareMathSymbol{\Psi}{\mathalpha}{letters}{"09}
\DeclareMathSymbol{\Omega}{\mathalpha}{letters}{"0A}
\makeatother

\def\footnoterule{\kern 20pt \hrule width 5pc \kern 2.6pt }
\usepackage{makeidx}  

\usepackage{amsmath,amsfonts,amssymb}
\usepackage{verbatim}
\usepackage{epsf}

\usepackage[bottom]{footmisc}
\usepackage{url}

\usepackage[pdftex]{graphicx}      

\usepackage{pdftricks}
\usepackage{pdfpages}
\DeclareGraphicsExtensions{.pdf}   
\usepackage[pdftex]{thumbpdf}      
\usepackage[pdftex,                
bookmarks=true,
bookmarksnumbered=true,
hypertexnames=false,
breaklinks=true,
linkbordercolor={0 0 1}]{hyperref} 
\hypersetup{
pdfauthor   = {J. Ratsaby},
pdftitle    = {On the Complexity of Binary Samples},
pdfsubject  = {Article},
pdfkeywords = {Binary functions, Density of sets, VC-dimension},
pdfcreator  = {LaTeX with hyperref package},
pdfproducer = {dvips + ps2pdf}
}
\newcommand{\real}{\mbox{\rm I\hspace{-0.2em}R}}
\newtheorem{theorem}{Theorem}
\newtheorem{remark}{Remark}

\newtheorem{claim}{Claim}
\newcommand{\Real}{ \real }
\def\mH{\mathcal{H}}
\def\mG{\mathcal{G}}
\def\sgn{\mbox{sgn}}
\def\mC{\mathcal{C}}
\def\I{\mathbb{I}}
\def\mH{\mathcal{H}}

\def\mF{\mathcal{F}}
\def\VC{\mbox{VC}}

\def\w{\omega}

\newcommand{{\integer}}{\mbox{\rm Z\hspace{-0.85em}Z}\,}
\newcommand{\be}{\begin{equation}}
\newcommand{\ee}{\end{equation}}

\def\squareforqed{\hbox{\rlap{$\sqcap$}$\sqcup$}}
\def\qed{\ifmmode\squareforqed\else{\unskip\nobreak\hfil
\penalty50\hskip1em\null\nobreak\hfil\squareforqed
\parfillskip=0pt\finalhyphendemerits=0\endgraf}\fi}

\def\blacksquareforqed{\hbox{$\blacksquare$}}
\def\bqed{\ifmmode\blacksquareforqed\else{\unskip\nobreak\hfil
\penalty50\hskip1em\null\nobreak\hfil\blacksquareforqed
\parfillskip=0pt\finalhyphendemerits=0\endgraf}\fi}

\usepackage{makeidx}  
\begin{document}
\bibliographystyle{plainnat}

\title{On the Complexity of Binary Samples}

\author{ Joel Ratsaby\\
{\small Electrical and Electronics Engineering Department}\\
{\small Ariel University Center of Samaria}\\
{\small ISRAEL}\\
 {\small\tt ratsaby@ariel.ac.il}}

\date{September 24, 2007}

\maketitle

\begin{abstract}
\vskip 0.3cm
Consider a class $\mH$ of  binary functions $h: X\rightarrow\{-1, +1\}$ 
on a finite interval $X=[0, B]\subset \Real$.
Define the {\em sample width} of $h$  on a finite subset (a sample) $S\subset X$
as
 \(
 \w_S(h) \equiv \min_{x\in S} |\w_h(x)|
 \)
where
$\w_h(x) = h(x) \max\{a\geq 0: h(z)=h(x), x-a\leq z\leq x+a\}$.
Let  $\mathbb{S}_\ell$ be the space of all samples in $X$ of cardinality $\ell$ 
and consider  sets  of wide samples, i.e., {\em hypersets} which are defined as
\(
A_{\beta, h} = \{S\in \mathbb{S}_\ell: \w_{S}(h) \geq \beta\}.
\)
Through an application of the Sauer-Shelah result on the density of sets
an upper estimate is obtained on the growth function (or trace) of 
the class  $\{A_{\beta, h}: h\in\mH\}$, $\beta>0$, i.e., on the number of possible
dichotomies obtained by intersecting
all hypersets with a fixed collection of samples $S\in\mathbb{S}_\ell$ of 
cardinality $m$.
The estimate is $2\sum_{i=0}^{2\lfloor B/(2\beta)\rfloor}{m-\ell\choose i}$.

\end{abstract}


\section{Overview}


Let $B>0$ and define the domain as $X = [0, B]$.
In this paper we consider the class $\mH$ of all {\em binary functions} 
 $h:X\rightarrow\{-1, +1\}$
which have
 only simple discontinuities, i.e.,  at any point $x$ the limits 
$h(x^+) \equiv\lim_{z\rightarrow x^+} h(z)$ from the right and similarly from the left $h(x^-)$
exist (but are not necessarily equal). 
A main theme of  our recent work  has been to characterize
binary functions based on their behavior on  a finite
subset of $X$.
In 
 \cite{AnthonyRatsaby-2006}
we showed that the problem of learning binary functions from a finite labeled sample
can  improve the generalization error-bounds
if  the learner obtains a hypothesis 
which in addition to minimizing the empirical sample-error  
is also `smooth' around  elements of the sample.
This  notion of smoothness (used also in  \cite{Ratsaby_jdmsc,RatsabyAADM})
is based on the simple notion of {\em width} of $h$ at $x$ which is defined as
\[
\w_h(x) = h(x) \max\{a\geq 0: h(z)=h(x), x-a\leq z\leq x+a\}.
%
\]
For a finite subset (also called {\em sample}) $S\subset X$
the {\em sample width} of $h$ denoted $\w_S(h)$
 is defined as 
 \[
 \w_S(h) \equiv \min_{x\in S} |\w_h(x)|.
 \]
This definition of width resembles the notion of sample {\em margin} of a real-valued function $f$ (see for instance
\cite{AB99}).
We say that a sample $S$ is  {\em wide} for $h$ if the width $\w_S(h)$ is large. 
Wide samples implicitly contain  more side information for instance 
about  a learning problem.
The current paper aims at estimating   the complexity of 
the class of wide samples for functions in  $\mH$. 
This complexity 
is related to a notion of  description complexity
 and knowing it 
enables to  compute the efficiency of information that is implicit in samples for learning
(see \cite{SOFSEM07}).

\section{Introduction}

For any logical expression $A$ 
 denote by $\I\{A\}$ the indicator function which takes the value $1$ or $0$ whenever 
the statement $A$ is true or false, respectively.
Let $\ell$ be any fixed positive integer and
define  the space $\mathbb{S}_\ell$  of all samples $S\subset X$ of size $\ell$.
On $\mathbb{S}_\ell$  consider  sets  of wide samples, i.e.,
\[
A_{\beta, h} = \{S\in \mathbb{S}_\ell: \w_{S}(h) \geq \beta\}, \quad \beta > 0.
\]
We refer to such sets as {\em hypersets}.
It will be convenient to associate with these sets the indicator functions on $\mathbb{S}_\ell$
 which are denoted as
\[
h'_{\beta, h}(S) = \mathbb{I}_{A_{\beta,h}}(S).
\]
These are referred to as {\em hyperconcepts} and
  we may write  $h'$ for brevity.
For any fixed width parameter $\gamma > 0$  define the {\em hyperclass}
\begin{equation}
\label{hprime}
\mathcal{H}'_\gamma = \left \{ h'_{\gamma,h}: h \in \mathcal{H}
\right \}.
\end{equation}
In words, $\mathcal{H}'_\gamma$ consists of all 
sets of subsets $S\subset X$ of cardinality $\ell$  on which the corresponding binary functions $h$
 are wide by at least $\gamma$.

The aim of the paper is to compute the complexity of the
 hyperclass $\mathcal{H}'_\gamma$ that corresponds to the class $\mH$.
Since the domain $X$ is infinite then so is $\mathcal{H}'_\gamma$ hence one cannot simply
measure its cardinality.
Instead  we  apply a standard combinatorial
measure of the complexity of a family of sets as follows:
suppose $Y$ is  a general domain and
 $\mG$ is  an infinite  class of subsets of $Y$. 
For  any subset $S=\{y_1, \ldots, y_n\}\subset Y$ let
\be
\label{Gm}
\Gamma_\mG(S) \equiv |\mG_{|S}|
\ee
where
 $\mG_{|S} = \{[\I_G(y_1), \ldots, \I_G(y_n)]: G\in \mG\}$. 
The  {\em growth function}  (see for instance \cite{AB99}) 
 is defined as
\[
\Gamma_{\mG}(n) = \max_{\{S: S\subset Y, |S|=n\}} \Gamma_\mG(S).
\]
It measures the rate in which the number of dichotomies obtained
by intersecting subsets $G$ of $\mG$ with a finite set $S$ 
increases as a function of the cardinality $n$ of $S$ in the maximal case (it is 
also called the trace of $\mG$ in \cite{Bollobas86}). 

Since we are interested in hypersets as opposed to simple sets $G$ (as above)
then we consider the trace
on a finite  collection $\zeta\subset \mathbb{S}_\ell$ of samples (instead of a finite sample $S$ as above).
It will be convenient to define  the cardinality of such a collection
 as the cardinality of the union of its component sets, i.e.,
for any given finite collection $\zeta \subset \mathbb{S}_\ell$ let
\be
\label{card}
|\zeta| = \left|\bigcup_{S: S\in\zeta} S\right|
\ee
and we use  $m$ to denote   a possible value of $|\zeta|$.
As a measure of complexity of  $\mH'_\gamma$ we  compute the growth as a function of $m$, i.e.
 \[
 \Gamma_{\mH'_\gamma}(m) = \max_{\zeta: \zeta \subset \mathbb{S}_\ell, |\zeta|=m}\Gamma_{\mH'_\gamma}(\zeta).
\]

\section{Main result}

Let us  state the  main result of the paper. 

\begin{theorem}
\label{th1}
Let $\ell, m>0$ be  finite integers and $B>0$ a finite real number. Let  $\mH$ be the class of  binary functions 
 on $[0,B]$ (with only simple discontinuities).
For a given width parameter value $\gamma > 0$, 
the corresponding hyperclass $\mH'_\gamma$ on the space $\mathbb{S}_\ell$  has a growth which is bounded as 
\[
\Gamma_{\mH'_\gamma}(m) \leq 
  2\sum_{i=0}^{2\lfloor B/(2\gamma)\rfloor}{m-\ell\choose i}.
\]
\end{theorem}
\begin{remark}
For $m > \ell + B/\gamma$, the following simpler bound  holds
\[
\Gamma_{\mH'_\gamma}(m) \leq  2\left(\frac{e\gamma(m-\ell)}{B} \right)^{\frac{B}{\gamma}}.
\]
\end{remark}

Before proving this result we need some additional notation.
We denote by $\langle a,b\rangle$ a generalized  interval set of the form
$[a,b]$, $(a,b)$, $[a,b)$ or $(a,b]$.
For a set $R$ we write $\I_R(x)$ to represent  the indicator function
of the statement $x\in R$. In case of an interval set
  $R=\langle a,b\rangle$ 
we write $\I\langle a,b\rangle$.

\begin{proof}
Any binary function $h$ may be represented by thresholding a real-valued
function $f$ on $X$, i.e., $h(x) = \sgn(f(x))$ where
for any $a\in\Real$, $\sgn(a)=+1$ or $-1$ if $a> 0$ or $a \leq 0$, respectively.
The idea is to choose a class $\mF$ of  real-valued functions $f$
which is  rich enough (it has to be infinite since there are infinitely many binary functions on $X$)
but is  as simple as we can find.
This is important since, as we will show, the growth function of $\mH'_\gamma$ 
is bounded from above
by the complexity of a class that is a variant of $\mF$.

We start by constructing such an $\mF$.
For a binary function $h$ on $X$
consider the corresponding set sequence  $\{R_i\}_{i=1,2,\ldots}$ which satisfies the following properties:
(a) $[0,B] = \bigcup_{i=1,2,\ldots}R_i$ and  for any $i\neq j$, $R_i\cap R_j=\emptyset$,
(b) $h$ alternates in sign over consecutive sets $R_i, R_{i+1}$,
(c) $R_i$ is an interval set  $\langle a, b\rangle$ with possibly $a=b$ (in which case $R_i=\{a\}$).
Hence $h$ has the following general form
\be
\label{hS}
h(x) = \pm\sum_{i=1,2,\ldots,}(-1)^i \I_{R_i}(x).
\ee
Thus there are exactly two functions $h$ corresponding uniquely to each sequence of sets $R_i$, $i=1,2,\ldots.$.
Unless explicitly specified, the end points of $X=[0, B]$
 are not considered roots of $h$, i.e., the default behavior is that outside
 $X$, i.e.,  $x < 0$ or $x > B$, the function `continues' with the same value it takes at the endpoint $h(0)$ or $h(B)$, respectively.
Now, associate with the set sequence $R_1, R_2, \ldots$ the unique non-decreasing sequence of right-endpoints
$a_1, a_2,\ldots$ 
which define these sets (the sequence may have up to two consecutive repetitions except for $0$ and $B$)
according to
\be
\label{intv}
R_i = \langle a_{i-1}, a_{i}\rangle, \; i=1,2,\ldots.
\ee
with the first left end point being $a_0=0$.
Note that different choices for  $\langle$ and $\rangle$ (see earlier definition of a generalized interval $\langle a, b\rangle$)
give different sets $R_i$ and hence different functions $h$.
For instance, suppose $X=[0,7]$ then 
the following set sequence  $R_1=[0, 2.4)$, $R_2=[2.4, 3.6)$, $R_3=[3.6, 3.6] = \{3.6\}$, 
$R_4=(3.6, 7]$ has a corresponding end-point sequence $a_1=2.4, a_2=3.6, a_3=3.6, a_4=7$.
 Note that 
a singleton set introduces a repeated value in this sequence.
As another example consider $R_1=[0,0]=\{0\}$, $R_2=(0, 4.1)$, $R_3=[4.1, 7]$ with 
$a_1 = 0$, $a_2=4.1$, $a_3=7$.

Next, define the corresponding sequence of midpoints
\[
\mu_{i} = \frac{a_{i}+a_{i+1}}{2}, \; i=1, 2, \ldots.
\]
Define the continuous real-valued function $f:X \rightarrow [-B, B]$ that corresponds to $h$ (via the
end-point sequence)  as follows:
\be
\label{fS}
f(x) = \pm\sum_{i=1,2,\ldots} (-1)^{i+1}(x-a_{i})\I[\mu_{i-1},\mu_i]
\ee
where we take $\mu_{0}=0$ (see for instance, Figure \ref{fig1}).
\begin{figure}[h]
\begin{center}
\epsfxsize=2.4in    \includegraphics[clip=true,scale = 0.7]{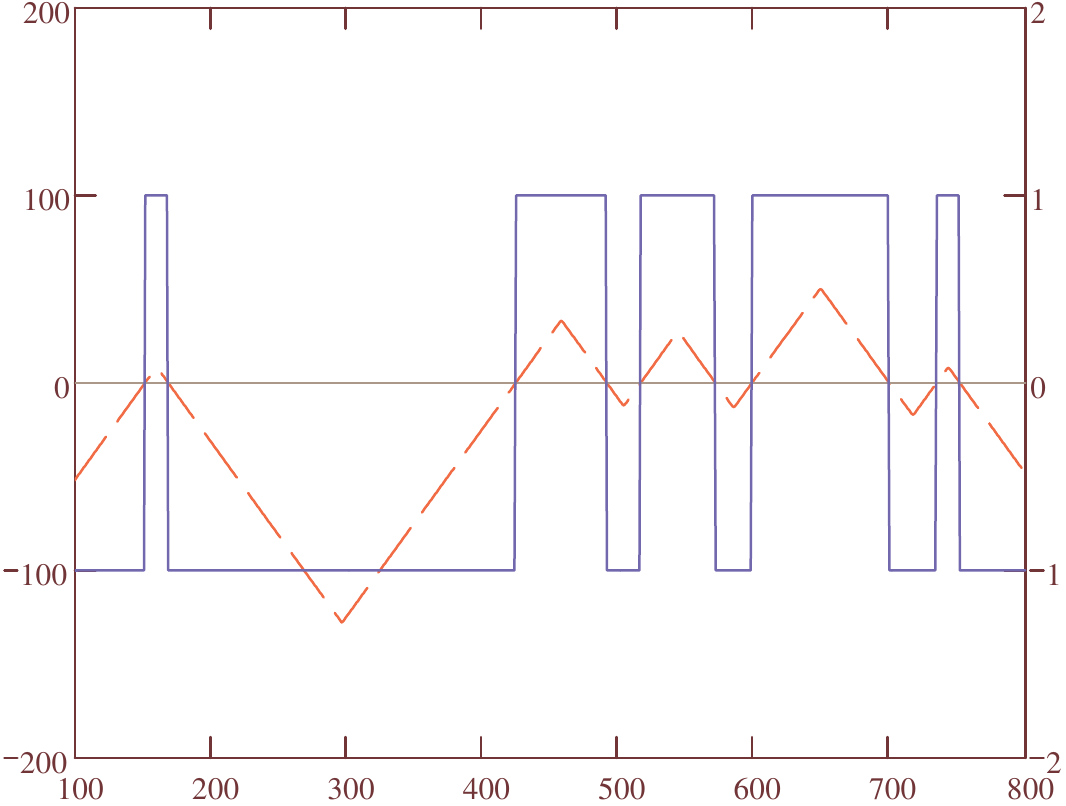} 
\end{center}
\caption{$h$ (solid) and its corresponding $f$ (dashed) on $X=[0, B]$ with $B=800$}
\label{fig1}
\end{figure}
Clearly, the value $f(x)$ equals the width $\w_h(x)$.
Note that for a fixed sequence of endpoints $a_i$, $i=1,2,\ldots$
 the function $f$ is invariant 
to the type of
 intervals  $R_i = \langle a_{i-1}, a_{i}\rangle$ that $h$ has,
for instance, the set sequence $[0, a_1)$, $[a_1, a_2)$, $[a_2, a_3]$, $(a_3, B]$
   and the sequence $[0, a_1]$, $(a_1, a_2]$, $(a_2, a_3]$, $(a_3, B]$ 
   yield different binary functions $h$ but the same width function $f$.
   For convenience, when $h$ has a finite number $n$ of interval sets $R_i$,
then the sum in (\ref{hS}) has an upper limit of $n$
and  we define $a_n=B$. Similarly, the sum in (\ref{fS}) goes up to ${n-1}$
and we define $\mu_{n-1} = B$.
Let us denote by 
\be
\label{fP}
\mF_+ = \{|f|: f\in \mF\}.
\ee
It follows that the hyperclass $\mH'_\gamma$ 
 may be represented in terms of the class $\mF_+$ as follows: define the hypersets
 \[
 A_{\beta, f} = \{S\in \mathbb{S}_\ell: f(x) \geq \beta, x\in S\}, \qquad \beta > 0, f\in \mF_+
 \]
with corresponding
 hyperconcepts $f'_{\gamma,f} = \mathbb{I}_{A_{\beta,f}}(S)$, let
 \[
 \mF'_\gamma = \{f'_{\gamma, f}: f\in \mF_+\}
 \]
 and 
 \be
 \label{let1}
 \mH'_\gamma = \mF'_\gamma.
 \ee
 Hence, it suffices to
  compute the growth function  $\Gamma_{\mF'_\gamma}(m)$.

Let us now begin to analyze the hyperclass $\mF'_\gamma$.
By definition, $\mathcal{F}'_\gamma$ is a class of indicator functions of subsets  of $\mathbb{S}_\ell$.
Denote by $\zeta_N\subset \mathbb{S}_\ell$ a collection of $N$ such subsets.
By a {\em generalized} collection we will mean a collection of subsets $S\subset X$ with  cardinality $|S|\leq \ell$.
Henceforth we fix a value $m$ and consider only collections
  \be
  \label{zn}
 \zeta_N, \text{ such that } |\zeta_N|=m
 \ee
where recall the definition of cardinality is according to (\ref{card}).
 Let us
 denote  the individual components of $\zeta_N$
by $S^{(j)} \in \mathbb{S}_\ell$,
 $1 \leq j \leq N$ hence
 \[
{\zeta}_N = \{S^{(1)}, \ldots, S^{(N)} \}.
\]
The growth function may be expressed as 
\begin{equation}
\label{thefirst}
\Gamma_{\mathcal{F}'_\gamma}(m) \equiv \max_{\zeta_N \subset \mathbb{S}_\ell, |\zeta_N|=m}
\Gamma_{\mathcal{F}'_\gamma}(\zeta_N) \equiv
\max_{\zeta_N \subset {\mathbb{S}_\ell}, |\zeta_N|=m} \left | \left\{[f'(S^{(1)}), \ldots, f'(S^{(N)})]: f'\in\mathcal{F}'_\gamma \right \}  \right |.
\end{equation}
Denote by $S^{(j)}_i$  the  $i^{th}$ element of the sample $S^{(j)}$ based on the ordering
of the elements of $S^{(j)}$ (which is induced by the ordering on $X$).
Then
\begin{eqnarray}
 \lefteqn{\Gamma_{\mathcal{F}'_\gamma}(\zeta_N)}\nonumber\\
  &=& \left | \left \{ \left[\mathbb{I}\left(\min_{x\in S^{(1)}} f(x) > \gamma \right) , \ldots,   
\mathbb{I}\left(\min_{x\in S^{(N)}}f(x) > \gamma \right)
 \right]: f\in \mF_+   \right \}    \right |\nonumber\\
 &=&
 \label{prod1}
 \left | \left \{ \left[\prod_{j=1}^\ell  \mathbb{I}\left(f(S^{(1)}_j) > \gamma \right) , \ldots,   
\prod_{j=1}^\ell  \mathbb{I}\left(f(S^{(N)}_j) > \gamma  \right) \right]:
 f\in \mF_+  \right \}    \right |.\qquad 
\end{eqnarray}
Order the elements  in each component  of $\zeta_N$ by the underlying ordering on $X$.
Then put the sets in lexical ordering starting with the first up to the $\ell^{th}$ element.
For instance, suppose $m=7$, $N=3$, $\ell=4$ and
 \[
 \zeta_3 = \{\ \{2, 8, 9, {10}\},  \{2, 5, 8, 9\}, \{3, 8, {10}, {13}\}\}
 \]
then the ordered version is
\[
\{\{2, 5, 8, 9\},  \{2, 8, 9, {10}\}, \{3, 8, {10}, {13}\}\}.
 \]
For any $x\in X$ let
\be
\label{tht}
\theta^\gamma_f(x) \equiv \mathbb{I}\left(f(x)> \gamma\right)
\ee
(we will sometimes write  $\theta_f(x)$ for short).
For any sample $S{^{(i)}}$ of cardinality $|S{^{(i)}}|\geq 1$ let
\[
e_{S{^{(i)}}}(f) = \prod_{j=1}^{|S{^{(i)}}|}  \theta_f(S^{(i)}_j).
\]
Then for  $\zeta_N$ we denote by
\[
v_{{\zeta}_{{N}}}(f) \equiv 
\left [e_{{S}^{(1)}}(f), \ldots, e_{{S}^{({N})}}(f)\right]
\]
where for brevity we sometimes write  $v(f)$.
Let
\[
V_{\mF_+}(\zeta_N) = \left \{v_{\zeta_N}(f): f \in \mF_+ \right \}
\]
or simply $V(\zeta_N)$.
Then from (\ref{prod1}) we have
\begin{equation}
\label{newEq}
\Gamma_{\mathcal{F}'_\gamma}(\zeta_N) = 
\left|{V_{\mF_+}}(\zeta_N)\right|.
\end{equation}
 Denote by $X'$ the union
 \be
 \label{xp}
 \bigcup_{j=1}^NS^{(j)} =  X' =\{x_i\}_{i=1}^m \subset X
 \ee
 and take  the elements to be  ordered
 as $x_i < x_{i+1}$, $1\leq i\leq m-1$.
The dependence of $X'$ on $\zeta_N$ is left implicit.
We will need the following procedure which maps  $\zeta_N$
to a generalized collection.

\mbox{}\\
\noindent {\bf Procedure G}:
{\em Given   $\zeta_N$ construct $\zeta_{\hat{N}}$ as follows:
Let $\hat{S}^{(1)} = S^{(1)}$. For any $2 \leq i \leq N$,  let
\[
\hat{S}^{(i)} = S^{(i)} \setminus \bigcup_{k=1}^{i-1} \hat{S}^{(k)}.
\]
Let  $\hat{N}$ be the number of non-empty sets $\hat{S}^{(i)}$.}
\mbox{}\\

Note that $\hat{N}$ may be smaller than $N$
since there may be an element of $\zeta_N$ which is contained in the union
of other elements of $\zeta_N$.
It is easy to verify by induction that the sets of ${\zeta}_{\hat{N}}$ are mutually exclusive
and  their union equals that of the original sets in $\zeta_N$.
We have the following:
\begin{claim}
\label{cl1}
\(
\left|{V_{\mathcal{F}_+}}(\zeta_N)\right|\leq \left|{V_{\mathcal{F}_+}}(G(\zeta_{{N}}))\right|.
\)
\end{claim}
\noindent{\em Proof}:
We make repetitive use of the following: let $A, B \subset X'$ be two non-empty sets and
let $C = B \setminus A$.
Then for any $f$, any $b\in \{0, 1\}$,
if $[e_{A}(f), e_{B}(f)] = [b, 0]$,   then  $[e_{{A}}(f), e_{C}(f)]$ may be either $[b, 0]$
or $[b, 1]$
since the elements in  $B$ which caused
the product $e_{B}(f)$ to be zero may or may not also be in ${C}$. 
In the other case
if $[e_{A}(f), e_{B}(f)] = [b, 1]$ then $[e_{{A}}(f), e_{{C}}(f)] = [b, 1]$.
Hence 
\[
\left| \left\{ \left [e_{A}(f), e_{B}(f)\right]:
 f\in \mathcal{F}_+
\right \}\right|
 \leq 
\left| \left\{
 \left[e_{{A}}(f), e_{{C}}(f) \right]:
  f\in \mathcal{F}_+
\right \} \right|.
\]
The same argument holds also for multiple $A_1, \ldots, A_k$, $B$ and $C = B \setminus \bigcup_{i=1}^k A_i$.
 Let ${\zeta}_{\hat{N}} = G(\zeta_N)$.
We now apply this to the following:
\begin{eqnarray}
\lefteqn{\left| \left\{ \left [e_{S^{(1)}}(f), e_{S^{(2)}}(f), e_{S^{(3)}}(f), \ldots, e_{S^{(N)}}(f)\right]:
 f\in \mathcal{F}_+
\right \}\right|}\nonumber\\
\label{e1}
&=&
\left| \left\{ \left [e_{\hat{S}^{(1)}}(f), e_{S^{(2)}}(f), e_{S^{(3)}}(f), \ldots, e_{S^{(N)}}(f)\right]:
 f\in \mathcal{F}_+
\right \}\right|\\
\label{e2}
&\leq&
\left| \left\{ \left [e_{\hat{S}^{(1)}}(f), e_{\hat{S}^{(2)}}(f), e_{S^{(3)}}(f),  \ldots, e_{S^{(N)}}(f)\right]:
 f\in \mathcal{F}_+
\right \}\right|\\
\label{e3}
&\leq&
\label{e4}
\left| \left\{ \left [e_{\hat{S}^{(1)}}(f), e_{\hat{S}^{(2)}}(f), e_{\hat{S}^{(3)}}(f), e_{{S}^{(4)}}(f)  \ldots, e_{S^{(N)}}(f)\right]:
 f\in \mathcal{F}_+
\right \}\right|\\
&\leq& \cdots\nonumber\\
&\leq & 
\label{eso}
\left| \left\{ \left [e_{\hat{S}^{(1)}}(h), e_{\hat{S}^{(2)}}(h), e_{\hat{S}^{(3)}}(h), e_{\hat{S}^{(4)}}(h),  \ldots, e_{\hat{S}^{(N)}}(h)\right]:
 f\in \mathcal{F}_+
\right \}\right|
\end{eqnarray}
where (\ref{e1}) follows since using $G$ we have $\hat{S}^{(1)} \equiv {S}^{(1)}$,
(\ref{e2}) follows by applying the above with $A = \hat{S}^{(1)}$, $B=S^{(2)}$ and
$C=\hat{S}^{(2)}$, (\ref{e3}) follows by letting $A_1=\hat{S}^{(1)}$, 
 $A_2=\hat{S}^{(2)}$, $B={S}^{(3)}$, and $C=\hat{S}^{(3)}$.
Finally, removing those sets $\hat{S}^{(i)}$ which are possibly empty 
leaves 
$\hat{N}$-dimensional vectors consisting only of the non-empty sets so
(\ref{eso}) becomes 
\(
\left| \left\{ \left [e_{\hat{S}^{(1)}}(f), \ldots, e_{\hat{S}^{(\hat{N})}}(f)\right]:
 f\in \mathcal{F}_+
\right \}\right|.
\)
\qed

Hence (\ref{prod1}) is bounded from above as
\begin{equation}
\label{es1}
\Gamma_{\mathcal{F}'_\gamma}(\zeta_N) \leq 
\left|{V_{\mathcal{F}_+}}(G(\zeta_{{N}}))\right|.
 \end{equation}
Denote by $N^* \equiv m-\ell+1$ and
define the following procedure which maps a generalized collection of sets in $X$ to another.

\mbox{}\\
\noindent{\bf Procedure Q}:
{\em Given a generalized collection $\zeta_{N} = \{S^{(i)}\}_{i=1}^N$, $S^{(i)}\subset X$. Construct   $\zeta_{N^*}$ 
as follows: 
let $Y = \bigcup_{i=2}^N  S^{(i)}$ and let the elements in $Y$
be ordered according to their ordering on $X'$ (we will refer to them as $y_1$, $y_2$, $\ldots$).
Let $S^{*(1)} = S^{(1)}$. 
For $2\leq i \leq m-\ell+1$,
 let $S^{*(i)} = \{y_{i-1}\}$.
}
\mbox{}\\
We now have the following:

\begin{claim}
\label{cl2}
For any $\zeta_N \subset \mathbb{S}_\ell$ with $|\zeta_N|=m$, then
\[
\left|{V_{\mF_+}}(G(\zeta_N))\right| \leq \left|{V_{\mF_+}}(Q(G(\zeta_N)))\right|.
\]
\end{claim}

\noindent{\em Proof}: 
Let $\zeta_{\tilde{N}} \equiv Q(G(\zeta_{N}))$ and as before $\zeta_{\hat{N}} = G(\zeta_N)$.
Note that by definition of Procedure $Q$, it follows that $\zeta_{\tilde{N}}$
consists of  $\tilde{N} = N^*$ non-overlapping sets,  the first $\tilde{S}^{(1)}$
having  cardinality $\ell$ and 
 $\tilde{S}^{(i)}$, $2\leq i \leq \tilde{N}$, each having 
  a single distinct element of $X'$.
Their union satisfies $\bigcup_{i=1}^{\tilde{N}}\tilde{S}^{(i)} = X'$. 

Consider the sets $V_{\mF_+}(\zeta_{\hat{N}})$, $V_{\mF_+}(\zeta_{\tilde{N}})$  and denote them
simply
by $\hat{V}$ and $\tilde{V}$.
For any $\hat{v}\in \hat{V}$ consider the following subset of $\mF_+$,
\[
B(\hat{v}) = \left\{f\in \mF_+: \hat{v}(f) = \hat{v}
 \right\}.
\]
We consider two types of $\hat{v}\in \hat{V}$.
The first does {\em not} have
 the following property:
 there exist functions ${f_\alpha}$, ${f_\beta}\in B(\hat{v})$
 with $\theta^\gamma_{f_\alpha}(x) \neq \theta^\gamma_{f_\beta}(x)$ for at least one element $x\in X'$. 
 Denote
by \(
\theta^\gamma_f \equiv [\theta^\gamma_f(x_1), \ldots, \theta^\gamma_f(x_m)].
\)
 Then in this case all 
 $f\in B(\hat{v})$
have the same $\theta^\gamma_f  = \hat{\theta}$, where $\hat\theta \in \{0,1\}^{m}$.
This implies that
\[
e_{\tilde{S}^{(1)}}(f) = e_{\hat{S}^{(1)}}(f) = \hat{v}_1
 \]
while for $2\leq j\leq \tilde{N}$ 
we have
\[
e_{\tilde{S}^{(j)}}(f) = \hat{\theta}_{k(j)}
\]
where $k:[N^*]\rightarrow[m]$ maps from the index of a (singleton) set $\tilde{S}^{(j)}$ to the index of an element
of $X'$  
and $\hat{\theta}_{k(j)}$ denotes the $k(j)^{th}$ component of $\hat{\theta}$.
Hence it  follows that  \[
|V_{B(\hat{v})}(\zeta_{\tilde{N}})| = |V_{B(\hat{v})}(\zeta_{\hat{N}})| .
\]
Let the second type of $\hat{v}$ satisfy  the complement condition, namely, 
 there exist functions ${f_\alpha}$, ${f_\beta}\in B(\hat{v})$
 with $\theta^\gamma_{f_\alpha}(x) \neq \theta^\gamma_{f_\beta}(x)$ for at least one point $x\in X'$.
If such $x$ is an element of  $\hat{S}^{(1)}$ 
 then the first part of the  argument  above holds
   and we still have
 \[
|V_{B(\hat{v})}(\zeta_{\tilde{N}})| = |V_{B(\hat{v})}(\zeta_{\hat{N}})|.
\]
If however there is also such an $x$  in some set $\hat{S}^{(j)}$, $2\leq j\leq \hat{N}$ then  since
the sets $\tilde{S}^{(i)}$, $2\leq i\leq \tilde{N}$ are  singletons then
there  exists some  $\tilde{S}^{(i)}\subseteq \hat{S}^{(j)}$ with
\[
e_{\tilde{S}^{(i)}}(f_{\alpha}) \neq e_{\tilde{S}^{(i)}}(f_{\beta}).
\]
Hence for this second type of $\hat{v}$ we have
\begin{equation}
\label{combine}
|V_{B(\hat{v})}(\zeta_{\tilde{N}})| \geq |V_{B(\hat{v})}(\zeta_{\hat{N}})| .
\end{equation}
Combining the above, then (\ref{combine}) holds for any $\hat{v}\in\hat{V}$.

Now, consider any two distinct $\hat{v}_\alpha$,  $\hat{v}_\beta \in \hat{V}$.
Clearly, $B(\hat{v}_\alpha) \bigcap B(\hat{v}_\beta) = \emptyset$  since
every $f$ has a unique $\hat{v}(f)$.
Moreover, for any
 $f_{a}\in B(\hat{v}_\alpha)$ and $f_b\in B(\hat{v}_\beta)$
 we have $\tilde{v}(f_a) \neq \tilde{v}(f_b)$ for the following reason: there must exist
 some set $\hat{S}^{(i)}$ and   a point $x\in \hat{S}^{(i)}$
 %
 %
 such that $\theta^\gamma_{f_a}(x) \neq \theta^\gamma_{f_b}(x)$ (since $\hat{v}_\alpha \neq \hat{v}_\beta$).
 If $i=1$ then they must differ on $\tilde{S}^{(1)}$, i.e., $e_{\tilde{S}^{(1)}}(f_\alpha) \neq 
 e_{\tilde{S}^{(1)}}(f_\beta)$.
  If  $2\leq i\leq \hat{N}$, then such an $x$
    is in some set $\tilde{S}^{(j)}\subseteq \hat{S}^{(i)}$ where $2\leq j\leq \tilde{N}$
  and therefore 
  $e_{\tilde{S}^{(j)}}(f_\alpha) \neq 
 e_{\tilde{S}^{(j)}}(f_\beta)$.
  Hence 
 no two distinct $\hat{v}_\alpha$, $\hat{v}_\beta$ map to the same $\tilde{v}$.
  We therefore have
 \begin{eqnarray}
 \left|{V_{\mF_+} }(\zeta_{\hat{N}})\right| &=& \sum_{\hat{v}\in\hat{V}}|V_{B(\hat{v})}(\zeta_{\hat{N}})| \nonumber\\
 \label{by8}
 &\leq& \sum_{\hat{v}\in\hat{V}}|V_{B(\hat{v})}(\zeta_{\tilde{N}})| \\
   &=&|V_{\mF_+}(\zeta_{\tilde{N}})| \nonumber
 \end{eqnarray}
 where (\ref{by8}) follows from (\ref{combine})
 which proves the claim.
\qed

Note that by construction of Procedure $Q$,  the dimensionality of the elements
of $V_{\mF_+}(Q(G(\zeta_N)))$ is $N^*$, i.e., $m-\ell+1$,
which holds for any   $\zeta_N$ (even maximally overlapping) and $X'$ as defined in (\ref{zn}) and (\ref{xp}).
Let us denote by $\zeta_{N^*}$ any set obtained by applying Procedure $G$  on any collection $\zeta_N$ followed by Procedure $Q$, i.e.,
\[
\zeta_{N^*} \equiv \left\{S^{*(1)}, S^{*(2)}, \ldots,  S^{*(N^*)}\right\}
\]
with a set $S^{*(1)}\subset X'$  of cardinality $\ell$
 and
\[
{S}^{*(k)} = \{x_{i_k}\}, \mbox{ where } x_{i_k} \in X'\setminus{S}^{*(1)}, \quad k = 2, \ldots, N^*.
\]
Hence we have
\begin{eqnarray}
\label{starIneq1}
\max_{\zeta_N\subset\mathbb{S}_\ell, 
 |\zeta_N|=m
}
 \Gamma_{\mathcal{F}'_\gamma}(\zeta_N) &\leq  &
\max_{\zeta_N\subset\mathbb{S}_\ell, 
 |\zeta_N|=m
} 
\left| V_{\mF_+}\left(Q(G(\zeta_{N}))\right)  \right|\\
\label{starIneq2}
&\leq & \max_{\zeta_{N^*}: |\zeta_{N^*}|=m
} \left| V_{\mF_+}(\zeta_{N^*})  \right|
\end{eqnarray}
where (\ref{starIneq1}) follows from (\ref{prod1}), (\ref{newEq}) and  Claims \ref{cl1} and \ref{cl2} while
 (\ref{starIneq2}) follows by definition of $\zeta_{N^*}$.
Now,  
\begin{eqnarray}
\left| V_{\mF_+}(\zeta_{N^*})  \right| &=&
\left| \left\{[e_{S^{*(1)}}(f), \ldots, e_{S^{*(N^*)}}(f)]: f\in \mF_+ \right\} \right|\nonumber\\
&\leq & 
\label{2ineq}
2\left| \left\{[e_{S^{*(2)}}(f), \ldots, e_{S^{*(N^*)}}(f)]: f\in \mF_+ \right\} \right|
\end{eqnarray}
where (\ref{2ineq}) follows trivially since $e_{S^{*(1)}}(f)$ is  binary.
So from (\ref{starIneq2}) we have 
\begin{eqnarray}
\lefteqn{
\max_{\zeta_N\subset\mathbb{S},
|\zeta_N|=m
}
 \Gamma_{\mathcal{F}'_\gamma}(\zeta_N) \leq  
2\max_{\zeta_{N^*}: |\zeta_{N^*}|=m
} 
\left| \left\{[e_{S^{*(2)}}(f), \ldots, e_{S^{*(N^*)}}(f)]: f\in \mF_+ \right\} \right|}\nonumber\\
\label{star3}
&\qquad\leq & 2\max_{x_1, \ldots, x_{m-\ell}\in X} 
\left| \left\{[\theta^\gamma_{f}(x_1), \ldots, \theta^\gamma_{f}(x_{m-\ell})]: f\in \mF_+ \right\} \right|\quad
\end{eqnarray}
where $x_1, \ldots, x_{m-\ell}$ run over any $m-\ell$ points in $X$.
Define the following infinite class of binary functions on $X$ by
\[
\Theta^\gamma_{\mF_+} = \{\theta^\gamma_f(x): f\in \mF_+\}
\]
and 
for any finite subset \[
X'' = \{x_1, \ldots, x_{m-\ell}\} \subset X
\]
let
\[
\theta^\gamma_f(X'')  = \left[\theta^\gamma_f(x_1), \ldots, \theta^\gamma_f(x_{m-\ell}) \right]
\]
and
\[
\Theta^\gamma_{\mF_+}(X'') = \{\theta^\gamma_f(X''): f\in \mF_+\}.
\]
We proceed to bound $|\Theta^\gamma_{\mF_+}(X'')|$.

The class $\Theta^\gamma_{\mF_+}$ is in one-to-one  correspondence with a class $\mC^\gamma_{\mF_+}$
 of sets $C_f\subset X$ which are defined as 
 \[
 C_f = \{x: \theta^\gamma_f(x) = 1\}, \quad f\in \mF_+.
 \]
We claim that any such set $C_f$ equals the union of at most $K \equiv \lfloor B/(2\gamma) \rfloor$
intervals.
To see this, note that based on the general form of $f\in \mF_+$ (see (\ref{fS}) and (\ref{fP})) 
in order for $f(x) > \gamma$  for every $x$ in an interval set $\frak{I}\subset X$ then
$\frak{I}$ must be contained in an interval set of the form  (\ref{intv})
and of length at least $2\gamma$.
 Hence for any $f\in\mF_+$ the corresponding set $C_f$ is comprised
 of no more than $K$ distinct intervals as $\frak{I}$.
 Hence the class $\mC^\gamma_{\mF_+}$ is a subset of the class $\mC_K$ of all sets 
 that are comprised of the union of at most $K$ subsets of $X$.
    A class $H$ is said to {\em shatter} $A$ if \(
\left|\{h_{|A}: h\in H\}\right| = 2^k.
\)
 The Vapnik-Chervonenkis dimension of $H$, denoted
as $VC(\mH)$, is defined as the cardinality of the largest set  shattered by $\mH$.
It is easy to show that the VC-dimension
   of $\mC_K$ is $\VC(\mC_K) = 2K$.
 Hence it follows from the Sauer-Shelah  lemma (see \cite{Sauer72})
 that the growth  of $\mC^\gamma_{\mF_+}$ on any finite set $X''\subset X$ of cardinality $m-\ell$ (see (\ref{Gm}))   satisfies 
 \[
 \Gamma_{\mC^\gamma_{\mF_+}}(X'') \leq  \sum_{i=0}^{2K}{m-\ell\choose i}.
 \]
Since 
\(
|\Theta^\gamma_{{\mF_+}}(X'')| = \Gamma_{\mC^\gamma_{\mF_+}}(X'') 
\)
then  from 
(\ref{let1}) and (\ref{star3}) it follows that
\[
|\Gamma_{\mH'_\gamma}(m)| \leq  2\sum_{i=0}^{2\lfloor B/(2\gamma)\rfloor}{m-\ell\choose i}
 \]
 which proves the statement of the theorem.
\end{proof}


\bibliography{jer}
\end{document}